\documentclass[reprint,amsmath,amssymb,pra,superscriptaddress,twocolumn]{revtex4-2}
\usepackage{import}
\usepackage{preamble}
\begin{document}
\title{Drag in Bose-Fermi Mixtures}
\author{Kai Yen Jee}
 \email{kj239@cornell.edu}
\author{Erich Mueller}
 \email{erich.mueller@cornell.edu}
\affiliation{Laboratory of Atomic and Solid State Physics, Cornell University, Ithaca, New York 14853, USA}%
\date{\today}

\begin{abstract}
We use kinetic theory to model the dynamics of a small Bose condensed cloud of heavy particles moving through a larger degenerate Fermi gas of light particles.  Varying the Bose-Fermi interaction, we find a crossover between bulk and surface dominated regimes -- where scattering occurs throughout the Bose cloud, or solely on the surface.  We calculate the damping and frequency shift of the dipole mode in a harmonic trap as a function of the magnetic field controlling an inter-species Feshbach resonance.
We find excellent agreement between our stochastic model and the experimental studies of Cs-Li mixtures.
\end{abstract}

\maketitle

\section{\label{intro}Introduction}

Many of the largest outstanding challenges in quantum matter lie in non-equilibrium dynamics.  Cold atom experiments \cite{nonequreview,blochreview,Navon167,chien}, and the associated theories \cite{ETH,Polkovnikov,eisertreview,jarzynski,turbulence}, have been instrumental in recent progress.   One important theme, explored in an experiment by the Chicago cold atom group \cite{chicagopaper}, is how energy is transferred from coherent motion into heat.  There they created interpenetrating clouds of quantum degenerate bosons and fermions, studying the dissipation which occurs when the clouds move relative to one-another.  Here we model that drag:  For repulsive Bose-Fermi interactions we find a crossover between surface-dominated and bulk-dominated scattering. For attractive Bose-Fermi interactions we find that individual fermions can spend substantial time inside the Bose cloud, leading to enhanced scattering.  We also model the dispersive forces, calculating how the Bose-Fermi interactions influence the dipole mode frequencies in a harmonic trap.

In the experiment, a small cloud of bosonic Cesium-133 sits within a larger gas of fermionic Lithium-6.  They are both trapped in a highly anisotropic ``cigar shaped'' optical trap, with an aspect ratio of roughly 10, but due to their different polarizabilites, the fermions experience a trap with an oscillation frequency that is roughly 5 times higher than the bosons. By using a Feshbach resonance \cite{feshbach}, the experimentalists control both the Cs-Cs and Cs-Li scattering lengths.  Due to quantum statistics, and the short-range nature of the interaction potentials, the Li atoms do not interact with one-another.  The main role of the Cs-Cs interactions is to set the density of the bosonic cloud.  When the Cs-Li scattering length, $a_{BF}$, is small, the two clouds interpenetrate, and the drag force is proportional to the overlapping volume, boson density, and the square of the scattering length.  On the other hand, when $a_{BF}$ is large and positive, fermions cannot penetrate the bosonic cloud.  In that regime the drag force is independent of both the scattering length and boson density, but is proportional to the surface area of the boundary. Attractive Bose-Fermi interactions leads to a novel regime where the fermions become trapped for a longer time in the boson cloud, leading to enhanced scattering effects. We calculate the drag force throughout these crossovers, capturing all of the structures seen in the experiment.

In addition to damping, the experimentalists observe a shift in the dipole-mode frequency of the bosonic cloud.  We argue that this shift is due to
buoyancy forces.  We precede our discussion of dissipation by first modelling these buoyancy forces in terms of the potential felt by fermions displaced from the bosonic cloud.  An equivalent model of these forces is given in Appendix~\ref{montecarlofrequency}, where
the buoyancy forces come from the ``lensing'' of fermion trajectories by the bosons.  This is analogous to the mechanism behind optical tweezers, and more closely parallels our treatment of the dissipative forces.  As would be expected, these two approaches  give identical numerical results, to within stochastic error.

In Section~\ref{buoyancy} we describe the buoyancy forces, and how they lead to a shift in the dipole mode frequency. In Sec.~\ref{kinetics}, we write down kinetic equations for the fermion atoms, and produce expressions for the momentum transfer from the bosons to the fermions. There we define a coefficient $\lambda$ which characterizes the drag force.  In Sec.~\ref{macrodynamics} we relate this microscopic quantity to the disipation of the dipole mode observed in the experiment.  Section~\ref{smalldrag} gives results in the limit where the Bose-Fermi scattering is weak.   Section~\ref{montecarlo} gives details of the Monte Carlo algorithm that we use for our numerical calculations.  Results are in Section~\ref{dragconstantresults}.
Section~\ref{conclusion} provides further discussion and conclusions. Three appendices follow.  The first describes how we self-consistently find the shapes of the boson and fermion clouds.  The second gives a technical argument regarding the weakly interacting limit.  The third gives our alternate model of the dispersive forces.

\section{Buoyancy}\label{buoyancy}
Archimedes' principle states that the buoyancy force on an object (in our case the Bose cloud) is equal and opposite to the external forces on the fluid it displaces (in our case the fermions). Due to the small size of our Bose cloud, this force is well approximated by
\begin{equation}
    F_{\rm buoyancy}\approx m_F \Delta N_F (\omega^x_F)^2 X_B.
\end{equation}
Here $X_B$ is the displacement of the boson cloud, $m_F$ is the mass of a single fermion, and $\omega^x_F$ is the harmonic trapping frequency of the fermions along the direction of the displacement.
We use a mean-field model to calculate the number of excess fermions  $\Delta N_F=\int n_F({\bf r})-n_
F^0({\bf r})\,d^3r$, where $n_F^0$ is the Fermi density in the absence of the bosons, and the integral is taken over the region occupied by the bosons. For sufficiently small velocities and displacements, $\Delta N_F$ can be taken from the equilibrium situation. Depending on the sign of the interactions, $\Delta N_F$ may be positive or negative.

\begin{figure}[tbp]
{\includegraphics[width=\columnwidth]{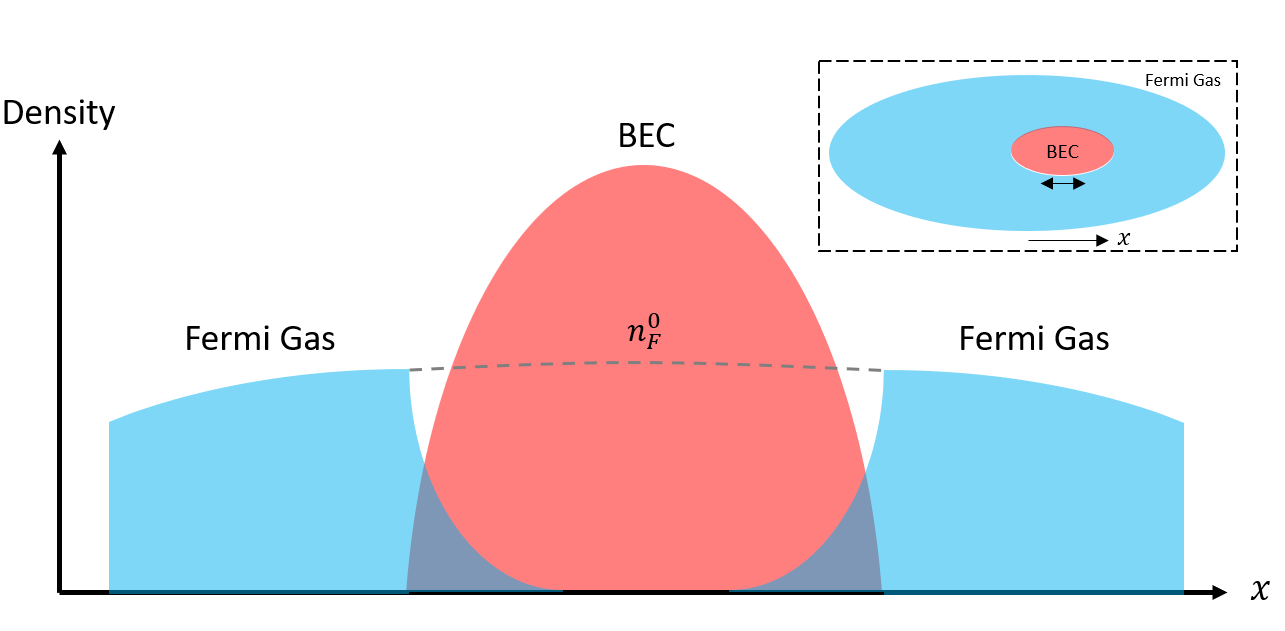}}
     \caption{(Color Online) Schematic of the density profiles of the harmonically trapped BEC (red) and Fermi gas (blue) for the case where the interspecies scattering length $a_{BF}$ is positive. The Fermi density in the absence of the Bose cloud, $n_F^0$, is shown as a dashed curve. Inset: Top-down view showing the BEC moving through the Fermi gas.
     }
     \label{buoyancydiagram}
 \end{figure}


Figure~\ref{buoyancydiagram} shows a schematic of the expected densities.
Within the Thomas-Fermi approximation, the fermion density profile is
\begin{equation}\label{fermiondensity}
    n_F = \frac{(2m_F)^{3/2}}{6\pi^2 \hbar^3}  \left[\mu_F -  V_F - g_{BF} n_B\right]^{3/2}. 
\end{equation}
Here, and in similar expressions in the rest of the paper, one should interpret $[x]^{3/2}$ as $\max(x,0)^{3/2}$. The Fermi trapping potential is $V_F=V_F(\textbf{r})=(1/2) m_F [(\omega_F^x)^2 x^2+(\omega_F^y)^2 y^2+
(\omega_F^z)^2 z^2]$, and $n_B=n_B({\bf r},t)$ is 
the equilibrium boson density.  The fermion chemical potential is $\mu_F$.  The coupling constant $g_{BF} =4\pi\hbar^2 a_{BF}/\mu$  is tuned via a Feshbach resonance.
The reduced mass is $\mu^{-1}=m_F^{-1}+m_B^{-1}$.
The equilibrium boson density is self-consistently found by numerically solving the Gross-Pitaevskii equation,
\begin{align}\label{bosondensity}
     \left(-\frac{\hbar^2 \nabla^2}{2m_B}  + V_B + g_{BB} n_B +  g_{BF} n_F\right)\psi = \mu_B \psi
\end{align}
with the Bose trapping potential being $V_B=V_B(\textbf{r})=(1/2) m_B [(\omega_B^x)^2 x^2+(\omega_B^y)^2 y^2+
(\omega_B^z)^2 z^2]$, and the boson chemical potential is $\mu_B$. The details of the procedure are in Appendix~\ref{densitysimulation}.  Similar analysis can be found in \cite{huang, capuzzi2004}.

We use the experimentally relevant values $\omega_B^x=2\pi \times 6.65$ Hz, $\omega_B^y=\omega_B^z=2\pi \times 118$ Hz,
$\omega_F^x=2\pi \times 34$ Hz, $\omega_F^y=\omega_F^y=2\pi \times 320$ Hz.  The chemical potentials are set by requiring that the total number of fermions and bosons are $N_F=20000$ and $N_B=30000$.  For the scattering lengths we take the s-wave Feshbach resonance curves to be $a_{BB}=1602.75 a_0 \left(1-(60.53/(B-820.37))\right)$ and $a_{BF}=-60 a_0 \left(2/(B-893) +1\right)$ respectively \cite{PhysRevA.87.010702}, where $a_0$ is the Bohr radius, and B is the applied external magnetic field in Gauss. We calculate profiles for $888<B<896$, roughly corresponding to the range of fields used in the experiments. 

We relate these density profiles to the frequency shift of the boson dipole mode by positing that the $\Delta N_F$ excess Fermions move with the bosons.  The equation of motion for the $x$-position of the boson cloud will then be
\begin{equation}
    (N_B m_B+\Delta N_F m_F) \ddot{X}_B = -(N_B m_B \omega_B^2  + \Delta N_F m_F \omega_F^2) X_B
\end{equation}
Assuming that $\Delta N_F m_F \ll N_B m_B$ and $\omega_F^2\ll \omega_B^2$ then to lowest order in these quantities, the shift in the dipole mode frequency is
\begin{equation}
    \delta\omega = \frac{1}{2} \frac{\Delta N_F m_F}{N_B m_B} \frac{\omega_F^2}{\omega_B}.
\end{equation}
The red dots in Fig.~\ref{freqshiftcomparison} show the resulting shift.  As described above and in Appendix~\ref{densitysimulation}, we find $\Delta N_F$ by self-consistently solving a Gross-Pitaevskii equation coupled with a Thomas-Fermi model for the fermions.  Numerical values for all the parameters are given above.

The small scattering length (ie. weak interaction) behavior can be understood analytically.
 To leading order, 
 \begin{equation}
    n_F 
    = n_F^0 \left(1-
    \frac{3 g_{BF} n_B}{2\mu_F}
    \right)
 \end{equation}
 where we have assumed that the Bose cloud is much smaller than the Fermi cloud.  Hence $\Delta N_F$ is proportional to $a_{BF}$ and the total number of bosons.  The resulting frequency shift is
 \begin{equation}\label{smallfreq}
     \delta\omega\approx -\left(\frac{\mu_F m_F^5}{8}    \right)^{1/2} \left(\frac{ (\omega_F^x)^2 }{\pi ^2  \hbar^3 m_B \omega_B^x}\right) g_{BF}
 \end{equation}
 This weak coupling result is shown as a solid red line in Fig.~\ref{freqshiftcomparison}. 
 
 The number of excluded fermions increases with $a_{BF}$, and hence the frequency shift becomes more negative as $a_{BF}$ increases.  The shift saturates at large positive $a_{BF}$ where the boson cloud excludes all fermions within their volume.  At negative scattering length, there is instead an accumulation of particles, and the frequency shift is predicted to be positive. 

  \begin{figure}[tbp]
{\includegraphics[width=\columnwidth]{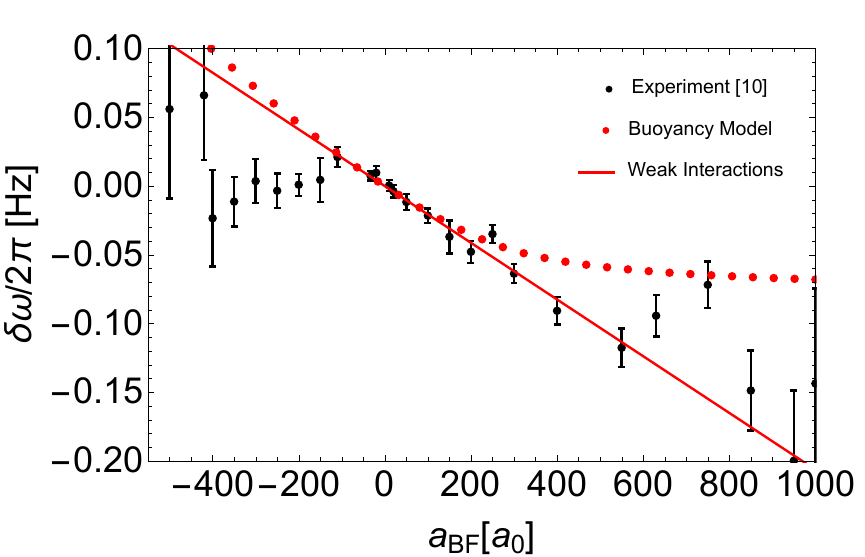}}
     \caption{(Color Online) Dipole mode frequency shift $\delta \omega$ 
     for a small Boson cloud inside of a larger Fermi gas in terms of the scattering length $a_{BF}$, corresponding to the experimental parameters in \cite{chicagopaper}. Small red dots show the Buoyancy model from Sec.~\ref{buoyancy}.  Black points with error bars show the experimental data.  The analytic weak-interaction expression, Eq.~(\ref{smallfreq}) is shown as a red line.}
     \label{freqshiftcomparison}
 \end{figure}
 
 While some of this structure is reproduced in the experimental data, there are notable differences. For instance, the experimental data approaches the weak-interaction limit at very high scattering lengths ($a_{BF}>800 a_0$) and very low scattering lengths ($a_{BF}< -500 a_0$).  We believe that these deviations are due to thermal effects:  Inelastic collisions lead to significant heating near the Feshbach resonance \cite{rudigrimm}, and 
 the experimentalists report strong deviations from the zero-temperature density profiles for these extreme scattering lengths.  At high temperatures, the boson density drops, and in the limit of low boson densities our theory predicts that the frequency-shift should approach the weak-coupling line.  We do not attempt a detailed modeling of the thermal profiles, as we do not have accurate estimates of the experimental temperature.
 
 One also sees deviations between the theory and experiment when $a_{BF}\sim 500 a_0$, and when 
 $a_{BF}\sim-300 a_0$.  These structures indicate that some extra physics is occurring in the experiments.  Possible ideas include: the excitation of a collective mode, or hydrodynamic effects like the generation of a wake or shockwave in the Fermi gas.  It is also possible that the excess fermions are not moving in lock-step with the bosons.  
 The source of these behaviors might be elucidated by carefully studying the in-situ density profiles.



\section{Drag}\label{drag}
\subsection{Setup} \label{setup}
Having modeled the dispersive forces in section~\ref{buoyancy}, we now turn to the central focus of this paper, modelling the dissipative forces.

We calculate the force that the fermions exert on the boson cloud by following the trajectories of individual fermions.  Since the bosons are much heavier than the fermions, their recoil can be neglected.  We add the impulses that each of the fermions experience, and use Newton's third law to deduce the force on the boson cloud.  For this calculation we neglect all external forces on the atoms, treating the fermion cloud as uniform, and taking the boson cloud to have the equilibrium shape calculated in section~\ref{buoyancy}.

As in the experiment, we take the boson velocity $\vec{\bf v}=v {\bf \hat x}$ to point in the ${\bf \hat x}$ direction, which is aligned with the long axis of the cloud.  
Under these circumstances, the net force on the bosons will be in the ${\bf \hat x}$ direction, ${\bf \vec{F}}=F {\bf \hat x}$.  
In Section~\ref{kinetics}
we calculate the coefficient of proportionality between force and velocity, $F=-\lambda v$.

\begin{figure}
\includegraphics[width=8cm]{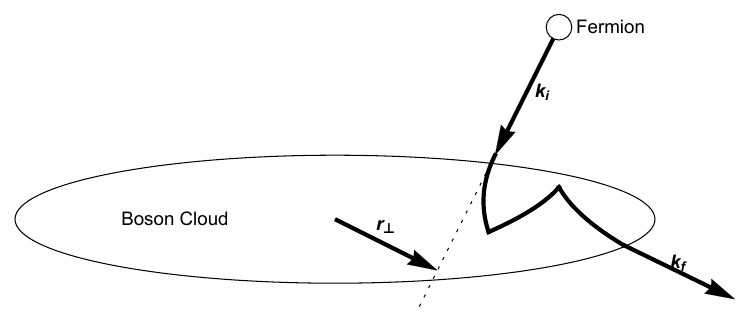}
\caption{\label{schematic}
Schematic depicting a trajectory of a fermion incident on the Boson cloud.  Two collision events are shown as kinks in the trajectory.  Vectors $\bf k_i, k_f$ and $\bf r_\perp$ are shown.
}
\end{figure}

\subsection{Kinetics} \label{kinetics}
We find it convenient to work in the frame where the bosons are stationary.  Thus we 
consider a stationary cloud of bosons with density $n_B(r)$ surrounded by a Fermi gas, whose center of mass is moving with velocity $v$. As illustrated in Fig.~\ref{schematic}, a single fermion of momentum $\hbar \bf k_i$ is incident on the bosonic cloud.  The fermion leaves with momentum $\hbar \bf k_f$, and the impulse imparted on the boson cloud is thus $\hbar(\bf k_i-k_f)$.  The total force is calculated by taking the total impulse imparted by all such collisions during time $\Delta t$, and dividing by $\Delta t$.


 As Fig.~\ref{schematic} shows, the trajectory is characterized by the incoming wave-vector and the ``impact parameter'' $\bf r_\perp$, which is the perpendicular displacement from the center of the cloud to the ray defining the path of the incoming fermion. Because of the nature of scattering, the motion of the fermion in the bosonic cloud is stochastic: There is some probability that the fermion scatters $0,1,2,\ldots$ times.  After each scattering event the fermion moves in a random direction.  Because the bosons are much heavier than the fermions, the magnitude of the fermion momentum (in the rest frame of the Bose cloud) is the same before and after scattering.  Between scatterings, the fermion moves in the mean-field potential from the bosons.
We define $P(\bm{k_i} \to \bm{k_f},r_\perp)$ as the probability density for a particle to leave with momentum $\bf k_f$, given that it entered with momentum $\bf k_i$, and impact parameter $\bf r_\perp$. 

In a time $\Delta t$, the number of fermions with momentum $\hbar {\bf k_i}$ that will enter the cloud, with impact parameter within cross-section $d^2 r_\perp$, is $\Delta N = f({\bf k_i}) \Delta V$. Here $f(\bm{k_i})$ is the fermion phase space density, and $\Delta V=\hbar |\bm{k_i}|\Delta t  \, d^2 r_\perp/m_f$ is the volume of space traced out by these particles during the time interval.  The mass of each fermion is $m_F$. Consequently the total impulse can be expressed as:
\begin{align}\label{kineticeqn}
    \bm{\Delta p} &= \int\!\! \frac{d^3\bm{k_i}}{(2\pi)^3} f(\bm{k_i}) \frac{\hbar |\bm{k_i}|}{m_F} \Delta t \int\!\! d^2 \bm{r_\perp} \int\!\! d^3\bm{k_f}\\\nonumber  &\qquad\qquad\qquad P(\bm{k_f} \to \bm{k_i},\bm{r_\perp})  \hbar (\bm{k_f}-\bm{k_i})
\end{align}
In equilibrium $f(\bm{k_i})$ is a Fermi function,
which, as long as $k_B T$ is small compared to the Fermi energy, can be modeled by its zero temperature form
%
\begin{align}\label{stepfunction}
    f(\bm{k_i}) &= \Theta\left(\left|\bm{k_i} - \frac{m_F\bm{v}}{\hbar}  \right|-k_{\rm fermi}\right)
\end{align}
where $\Theta(x)$ is the step function.
Linearizing for small $\bf v$, $f=f_0 +\delta f$
with
\begin{eqnarray}
    \delta f(\bm{k_i})&=&\frac{m_F\bm{v} \cdot \bm{k_i}}{\hbar|\bm{k_i}|} \delta(|\bm{k_i}|-k_{\rm fermi})
\end{eqnarray}
where $\delta(x)$ is the Dirac Delta function. By symmetry, Eq.~(\ref{kineticeqn}) vanishes if we replace $f$ with $f_0$.  Therefore to linear order,
%
   $ \bm{\Delta p} = -\lambda \bm{v} \Delta t
   $
with
\begin{align} \label{lambda}
    \lambda = -&\int \frac{d^3\bm{k_i}}{(2\pi)^3}  \delta(|\bm{k_i}|-k_{\rm fermi})(\hat{\bm{v}} \cdot \bm{k_i}) \int d^2 \bm{r_\perp} \nonumber\\
    &\qquad \qquad  \int d^3\bm{k_f} P(\bm{k_i} \to \bm{k_f},\bm{r_\perp})  \hbar (\bm{k_f}-\bm{k_i})\cdot \hat{\bm{v}}
\end{align} 
In section~\ref{montecarlo} we explain how to sample from $P$, and calculate $\lambda$ using a Monte-Carlo algorithm.  Section~\ref{smalldrag} analytically calculates this integral in the limit of weak Bose-Fermi interactions.
Section~\ref{macrodynamics} relates $\lambda$ to the macroscopic observables.

Interactions between the Bose and Fermi clouds play two roles here:  (1) The fermions feel a mean-field potential from the bosons, which causes their trajectories to curve.  These mean-field forces are proportional to $g_{BF}$.  (2) The fermions can experience hard-scattering events. The cross-section for these events are proportional to $g_{BF}^2$.  In our analytic treatment of the weak-interaction limit we are able to separately consider the contributions, but in our numerics in Sec.~\ref{montecarlo} we include both these effects together. This decomposition into mean-field and scattering terms is standard \cite{landauphysicalkinetics}. 
\section{Weak Interactions}\label{smalldrag}

Here we calculate $\lambda$ in the limit of small $a_{BF}$.  We will find that the leading behavior is
\begin{eqnarray}\label{lambda1}
\lambda &=&
\frac{2\hbar k_{\rm fermi}^4}{3\pi} N_B a_{BF}^2 
\end{eqnarray}

The lowest order contribution to the drag coefficient $\lambda$ comes from  scattering -- and is therefore proportional to the scattering cross-section $\sigma=4\pi a_{BF}^2$.  The result is proportional to $N_B$, as in this limit the probability of scattering off each boson is independent.  The dependence on $k_{\rm fermi}$ has two components: (1) The density of fermions is proportional to $k_{\rm fermi}^3$, and (2) their average velocity is proportional to $k_{\rm fermi}$.\\

In the following two subsections we derive Eqs. (\ref{lambda1}) by first
showing that the contributions from mean-field effects can be neglected:  Section~\ref{mfc} shows the linear in $a_{BF}$ terms vanish, and 
%
Appendix~\ref{higherorderlambda}, shows that the quadratic terms also vanish. Section~\ref{scat} calculates the leading order scattering contributions, which give  Eq. (\ref{lambda1}).

\subsection{Contributions from the Mean Field Potential}\label{mfc}

We consider the trajectory of a single fermion, defining $\boldsymbol{k}(t)$ to be its momentum as a function of time.  The position of the fermion is $\boldsymbol{r}(t)$. In the absense of scattering these obey:
\begin{align}\label{diffeq}
    \diff{\boldsymbol{k}}{t} &= -\frac{1}{\hbar} \nabla V(\boldsymbol{r}(t))\\
    \diff{\boldsymbol{r}}{t} &= \frac{\hbar \boldsymbol{k}(t)}{m_F}
\end{align}
where $V(\boldsymbol{r}(t)) = g_{BF} n_B (\boldsymbol{r}(t))$. 
We expand $ \boldsymbol{k}(t)$ and $ \boldsymbol{r}(t)$ in powers of $g_{BF}$ as:
\begin{align}
    \boldsymbol{k}(t) &=  \boldsymbol{k}_i + g_{BF} \boldsymbol{k}^{(1)}(t) + \frac{g_{BF}^2}{2} \boldsymbol{k}^{(2)}(t) + \hdots\\
    \boldsymbol{r}(t) &=  \boldsymbol{r}^{(0)}(t) + g_{BF} \boldsymbol{r}^{(1)}(t) + \frac{g_{BF}^2}{2} \boldsymbol{r}^{(2)}(t) + \hdots
\end{align}
which defines the approximants $\boldsymbol{k}^{(1)}$ and
$\boldsymbol{r}^{(1)}$. Integrating the zeroth order term yields
\begin{eqnarray}
\boldsymbol{r}^{(0)}(t) = \boldsymbol{r}_i+\hbar \boldsymbol{k}_i t/m_F.
\end{eqnarray}
To this order the momentum is a constant. Since the zeroth order path is a straight line, the final momentum 
can then be expressed as a geometric integral
\begin{eqnarray}
     g_{BF}\boldsymbol{k}^{(1)}(T)
    &=&
    -\int_0^T dt \, \frac{ g_{BF}}{\hbar}\nabla n_B\left(
    \boldsymbol{r}(t)
    \right)\\
    &=& -\int dr_\parallel \frac{m_F g_{BF} \nabla n_B({\bf r})}{\hbar^2 |\boldsymbol{k}_i |}\label{momchange}
\end{eqnarray}
where we have expressed the integral in terms of   $r_\parallel$, the distance fermion has moved along the direction of motion: ${\bf r}={\bf r}^{(0)}$ $={\bf r_\perp} + {\bf \hat k_i} r_\parallel$. 

When we substitute Eq.~(\ref{momchange}) into Eq.~(\ref{lambda}), the probability distribution becomes a delta-function, $P({\bf k_i}\to{\bf k_f},r_\perp)=\delta^3({\bf k_f} - {\bf k_i}+ g_{BF}\boldsymbol{k}^{(1)}(T))$, and the ${\bf k_f}$ integral is trivial.

For ease of notation, we define
\begin{align}
    \int_{\hspace{-2mm}\raisebox{-2mm}{\mbox{$\scriptstyle |\bm{k_i}|=k_{\rm fermi}$}}}\hspace{-1.1cm}d^2\bm{k_i} \equiv \int d^3\bm{k_i} \, \delta(|\bm{k_i}|-k_{\rm fermi})
\end{align}
so that we may write
\begin{align}\label{meanfieldlambda}
    \lambda = &
    \int_{\hspace{-2mm}\raisebox{-2.5mm}{\mbox{$\scriptstyle |\bm{k_i}|=k_{\rm fermi}$}}}\hspace{-1.1cm}\frac{d^2\bm{k_i}}{(2\pi)^3}  (\hat{\bm{v}} \cdot \bm{k_i}) \int\!\! d^2 \bm{r_{\!\perp}}\!   \int_0^L \!\!\!dr_\parallel \frac{m_F g_{BF} \,\hat{\bm{v}}\cdot\nabla n_B({\bf r}) }{\hbar |\boldsymbol{k}_i |} \nonumber\\
    = & \frac{m_F g_{BF}}{\hbar}\int_{\hspace{-2mm}\raisebox{-2.5mm}{\mbox{$\scriptstyle |\bm{k_i}|=k_{\rm fermi}$}}}\hspace{-1.1cm}\frac{d^2\bm{k_i}}{(2\pi)^3}  (\hat{\bm{v}} \cdot \hat{\bm{k}}_i) \int d^3 \bm{r}\,\hat{\bm{v}}\cdot \nabla n_B({\bf r}) 
\end{align}
In the second line we have combined the integrals over $\bm{r_\perp}$ and $r_\parallel $  
into a volume integral, which is independent of $k_i$. The integrand in Eq.~(\ref{meanfieldlambda}) is odd in $\bf k_i$, and hence the integral vanishes. Thus we see that there is no mean field contribution to $\lambda$ which is linear in $g_{BF}$.

\subsection{Contributions from Scattering}\label{scat}
Since there are no linear or quadratic contribution to $\lambda$ from the mean-field [see section~\ref{mfc} and appendix~\ref{higherorderlambda}]
we can neglect the mean field in calculating the leading order contribution from scattering.
Thus we treat the fermion trajectory as a sequence of straight-line paths between scattering events. In the weakly interacting limit, there will be at most one scattering event. The probability that a fermion will have a scattering event when it travels from position $\bf r$ to ${\bf r} +d{\bf r} $  is  $dP=\sigma n_B({\bf r}) dr$. For a given incoming wave-vector and transverse position, the total probability of a scattering will be:
\begin{align}
     P_{total} = \int d^3\bm{k_f} \,  P(\bm{k_i} \to \bm{k_f},r_\perp) = \int d r_\parallel \, \sigma n_B({\bf r}),
\end{align}
where, as before, $r_\parallel$ is the component of the position parallel to the incoming wave-vector, ${\bf r}={\bf r}^{(0)}$ $={\bf r_\perp} + {\bf \hat k_i} r_\parallel$.  The scattering event will be isotropic, with the direction of $\bf k_f$ uniformly distributed on a sphere of radius $k_{\rm fermi}$.  Thus, 
\begin{eqnarray}
\int d^3\bm{k_f} \,  P(\bm{k_i} \to \bm{k_f},r_\perp)
{\bf k_f}&=&0\\
\int d^3\bm{k_f} \,  P(\bm{k_i} \to \bm{k_f},r_\perp)
{\bf k_i}
&=& P_{total} {\bf k_i}.
\end{eqnarray}
Using these expressions in Eq.~(\ref{lambda}), we find
\begin{align}
     \lambda &= \hbar \int_{\hspace{-2mm}\raisebox{-2.5mm}{\mbox{$\scriptstyle |\bm{k_i}|=k_{\rm fermi}$}}}\hspace{-1.1cm}\frac{d^2\bm{k_i}}{(2\pi)^3} \,\hat{\bm{v}} \cdot \bm{k_i} \int d^2 \bm{r_\perp}   \int d r_\parallel \, \sigma n_B({\bf r}) \, \hat{\bm{v}}\cdot\bm{k_i} 
\end{align}
but $\int d^2 \bm{r_\perp}  \int dr_\parallel \, n_B = \int d^3 \bm{r} \, n_B = N_B$, the total number of bosons, and so we find 
\begin{align}
     \lambda &= \hbar \int_{\hspace{-2mm}\raisebox{-2.5mm}{\mbox{$\scriptstyle |\bm{k_i}|=k_{\rm fermi}$}}}\hspace{-1.1cm}\frac{d^2\bm{k_i}}{(2\pi)^3}  (\hat{\bm{v}} \cdot \bm{k_i})^2 \, \sigma N_B = \frac{\hbar \sigma N_B}{3}\int_{\hspace{-2mm}\raisebox{-2.5mm}{\mbox{$\scriptstyle |\bm{k_i}|=k_{\rm fermi}$}}}\hspace{-1.1cm}\frac{d^2\bm{k_i}}{(2\pi)^3}   |\bm{k_i}|^2 \, \nonumber \\
    &= \frac{\hbar k_{\rm fermi}^4 \sigma N_B}{6\pi^2 } =\frac{2\hbar k_{\rm fermi}^4  a_{BF}^2 N_B}{3\pi }
\end{align}
which is independent of the geometry of the boson cloud, as expected in this weak scattering regime. 

\section{Macroscopic Dynamics}\label{macrodynamics}

We now connect the damping coefficient $\lambda$ to the experimentally observed decay of the dipole mode.

Neglecting the buoyancy forces, which are small, 
The equation of motion of the center of mass of the boson cloud $X_B$, and the center of mass of the fermionic gas $X_F$ is:
\begin{align}\label{macro}
    \begin{pmatrix} M_B \ddot{X_B} \\ M_F \ddot{X_F} \end{pmatrix} =\begin{pmatrix} -M_B \omega_B^2 X_B 
    +F_{FB}
    \\-M_F \omega_F^2 X_F
    -F_{FB}
    \end{pmatrix},  
\end{align}
where $M_B=N_B m_B$ and $M_F=N_F m_F$ are the total masses of the bosons and fermions.  $F_{FB}$ is the force on the bosonic cloud from the fermionic cloud,
\begin{equation}
    F_{FB}
    =- \lambda (\dot{X_B} - \dot{X_F}).
\end{equation}
Equation~(\ref{macro}) is easily integrated.  In particular, for the experimentally relevant case where $\omega_F^x$ and $\omega_B^x$ are very different, the 
%
normal modes consist of either the bosons moving, with the fermions stationary, or the fermions moving with the bosons stationary.  These modes have frequencies near $\omega_F$ and $\omega_B$.  Their damping rates are 
\begin{equation}\label{macdamp}
\Gamma_F=\frac{\lambda}{2M_F},\qquad
\Gamma_B=\frac{\lambda}{2 M_B}.
\end{equation}
Furthermore, $N_B\approx N_F$ and $m_B/m_F= 133/6$, so $M_B\gg M_F$ and hence $\Gamma_F\gg\Gamma_B$.  The Fermion cloud's motion rapidly damps out. The experiment directly measures $\Gamma_B$, which is related to the microscopic coefficient $\lambda$ through Eq.~(\ref{macdamp}).

\section{Monte Carlo Simulation}\label{montecarlo}
To sample the fermion trajectories, we  convert Eq.~(\ref{lambda}) into a Monte Carlo sum. We write  $\bf k_i$ and $\bf k_f$ in spherical coordinates $(|\bf k_i|,\theta, \phi)$ and $(|\bf k_f|,\theta_{\rm out}, \phi_{\rm out})$.  We align the polar axis with
the cloud, which is also aligned with $\bf v$. Hence
$\hat{\bm{v}} \cdot \bm{k_i} =|\bm{k_i}| \cos\theta$.
Because of the cylindrical symmetry of the cloud, we can always choose $\phi=0$.
We denote the long-axis of the cloud as $\hat x$, and take ${\bf k_i}$ to lie in the $\bf \hat x-\hat y$ plane.
After straightforward simplification, we have
\begin{align} \label{lambda3}
   \lambda 
   &= \frac{\hbar k_{\rm fermi}^4}{(2\pi)^2}\int_{-1}^1 d(\cos\theta)\int d^2 \bm{r_\perp} \int d^3\bm{k_f}  P(\bm{k_i} \to \bm{k_f},\bm{r_\perp}) \nonumber \\
   &\qquad \qquad \qquad \qquad \qquad \qquad  \cos\theta(\cos\theta_{\rm out} -\cos\theta)
\end{align}
We parameterize the impact parameter ${\bf r_\perp}$ as
\begin{equation}
{\bf r_\perp}=
r_1 {\bf \hat z} + r_2 {\bf \hat k_\perp}
\end{equation}
where $\bf\hat k_\perp$ is a unit vector in the $\bf \hat x-\hat y$ plane which is perpendicular to $\bf k_i$.  We denote the radius of the Bose cloud in the $\bf\hat x$ direction as $a$, and its radius in the other two directions as $c$.  We numerically find $c$ and $a$ from our calculation in Sec.~\ref{buoyancy}.

We then calculate $\lambda$ in Eq.~(\ref{lambda3}) by randomly generating a set of trajectories. We first choose $\cos\theta$ and $r_1$  uniformly in $[-1,1]$ and $[-c,c]$ respectively.  We then choose $r_2$ uniformly in $[-\xi(\theta,r_1),\xi(\theta,r_1)]$,  where
\begin{equation}
\xi(\theta,r_1)=\frac{1}{c} \sqrt{c^2-r_1^2}
\sqrt{c^2\cos^2(\theta)+a^2\sin^2(\theta)}.
\end{equation}
These bounds are chosen to give the tightest rectangular aperture which fully contains the Bose cloud.
The drag coefficient is then
%
%
\begin{align}\label{lambdaexpression}
    \lambda=\frac{\hbar k_{\rm fermi}^4 }{2\pi^2  }  \frac{1}{N}\sum_{j=1}^N  A_{j}\cos\theta_j \, (\cos\theta_{out,j}-\cos\theta_j)
\end{align}
where $j$ denotes the sample index,  $N$ is the total number of samples, and $A_j=4 c \xi(\theta_j,r_{1,j})$ is the area of the aperture. Each sample is independent, so errors are simple to estimate.

To calculate the trajectory, and hence $\theta_{out,j}$, we use a temporal finite difference scheme.  We start with an initial $\bf r=r_0$ that is outside of the cloud, and set $\bf k_0=k_i$.  We choose a small timestep $\Delta t$.  In each time step we use the symplectic algorithm, updating ${\bf r}_{i+1}= {\bf r}_i+ (\hbar/m_f) \Delta t {\bf k}_i$ and ${\bf k}_{i+1}={\bf k}_i- \nabla V_{\rm mf}({\bf r}_{i+1})\Delta t/\hbar$.  We then calculate the probability of scattering during that time step,
$p=\sigma n_B({\bf r}_i) (\hbar/m_f) \Delta t {\bf k}_i$.  We generate a random number $s\in[0,1]$, and if $s<p$ we rotate ${\bf k}$ to point in a random direction.  We repeat until the fermion exits the cloud.

    
    
\begin{figure}
     \includegraphics[width=9cm]{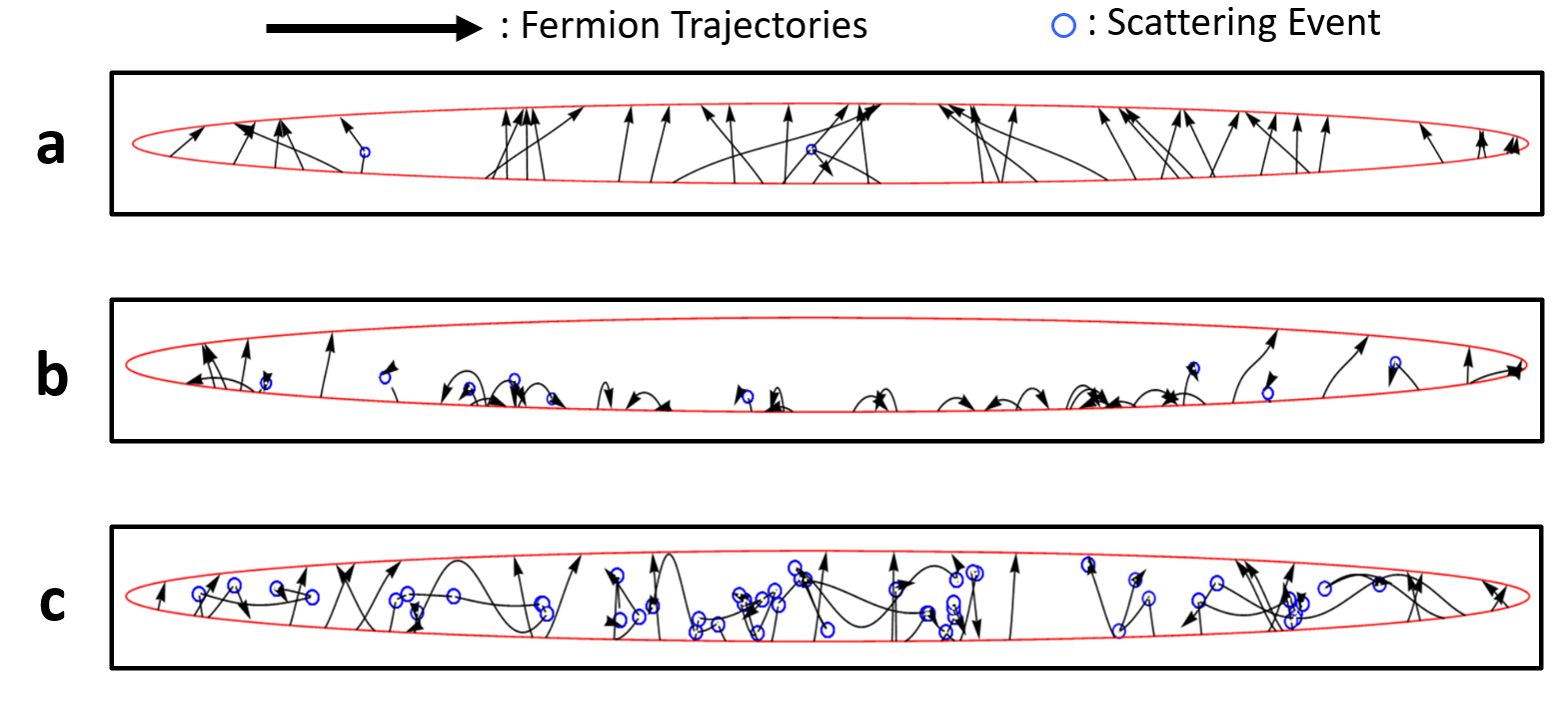}
     \caption{(Color Online) Representative fermion trajectories in different regimes: (a) small scattering lengths (Shown: $a_{BF} = 60 a_0$), (b) large positive scattering lengths (Shown: $a_{BF} = 340 a_0$), and (c) large negative scattering lengths (Shown: $a_{BF} = -340 a_0$) Red ellipsoids represent the Thomas-Fermi radii of the Boson cloud. Displayed trajectories begin in the $x$-$z$ plane with $k_y=0$ and positive $k_z$, and are projected to 2D for visualization. Scattering events are labeled with blue circles.}
     \label{trajectoryschematic}
 \end{figure}

Representative trajectories are shown in
Fig.~\ref{trajectoryschematic}. 
 For small scattering lengths, trajectories are nearly straight, with rare scattering events.  For large repulsive scattering lengths, the trajectories are highly curved, with the fermions unable to penetrate far into the Bose cloud.  Despite the large cross-sections, scattering events are relatively rare -- as the curved trajectories mean that the fermions spend very small amounts of time in the cloud.  For large attractive interactions, one occasionally sees spiral trajectories where the fermions spend large amounts of time in the Bose cloud.  These lead to many more scattering events.


\section{Drag Constant Results}\label{dragconstantresults}

\begin{figure}[hbt!]
{\includegraphics[width=\columnwidth]{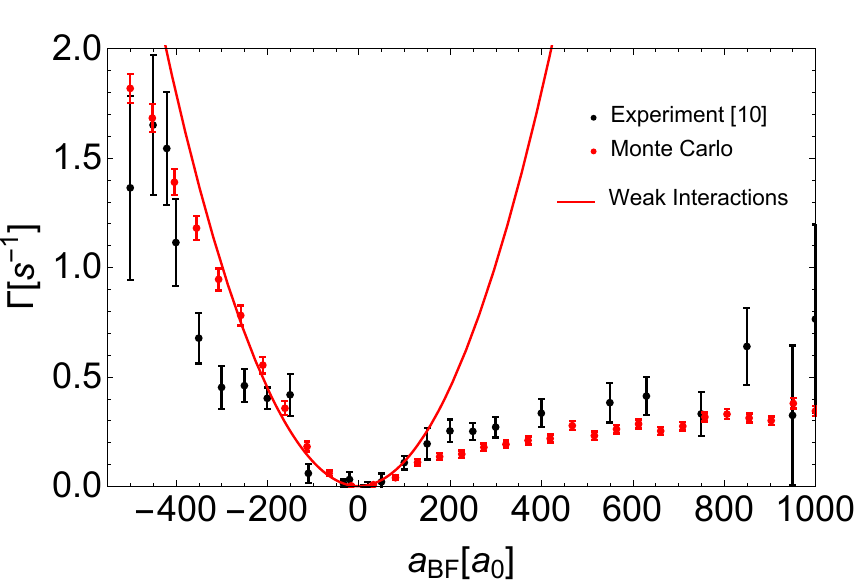}}
     \caption{(Color Online) Drag constant $\Gamma$ as a function of of the Bose-Fermi scattering length $a_{BF}$.  Red points with small error bars represent our Monte-Carlo calculations.  Black points with large error bars correspond to the experimental data from \cite{chicagopaper}.  The analytic weak-interaction limit is shown as a solid red line.   For the Monte-Carlo results, each point represents 10000 samples of fermion trajectories, with error bars corresponding to the 
     standard error of the mean.
     }
     \label{dragcomparison}
 \end{figure}

The results of the simulation from Section~\ref{montecarlo}, are shown in Fig.~\ref{dragcomparison}, along with the experimental results from \cite{chicagopaper}.  For small scattering lengths, the drag coefficient is quadratic in $a_{BF}$ -- agreeing with what one expects from our analytic weak coupling calculation.

At large positive scattering length, the drag constant saturates -- representing a crossover to a surface dominated regime. Full saturation is not achieved in this figure.  
For attractive interactions, the numerical data largely tracks the weak-coupling curve.  


There is good quantitative agreement between the experiment and our simulation, with the exception of the regime where $a_{BF}$ is between $-400$ and $-200 a_0$,  There the experimental data shows a pronounced plateau, which is not seen in our numerics.  There is a similar anomaly in the 
 the frequency shift data (Fig~\ref{freqshiftcomparison}). We therefore hypothesize that these two features may be related in some way.  Note: the other anomalies from Fig~\ref{freqshiftcomparison} do not appear to have counterparts in Fig.~\ref{dragcomparison}.

\section{Summary and Outlook} \label{conclusion}
We used a simple kinetic model to quantitatively explain the behavior of a Bose condensate of heavy atoms immersed in a larger cloud of quantum degenerate fermions.

We treated
the Bose condensate 
as a monolithic object, characterized by its center-of-mass position.  We described the fermions via a quantum Boltzmann equation, which leads to a 
fluid mechanics picture of the dynamics.  
For example, the Bose cloud experiences a buoyancy force, which we model using Archimedes principle.  The bosons also experience a drag force, which we calculate through a Monte Carlo algorithm that follows the trajectories of individual fermions.  There are natural parallels with classical models of Brownian motion. It is profoundly satisfying that these simple mechanical models are able to quantitatively describe dissipation in a Bose-Fermi mixture. We believe that such models can be used to describe other types of experiments involving Bose-Fermi mixtures \cite{PhysRevA.92.043604,PhysRevLett.119.233401,PhysRevLett.118.055301}, as well as other types of cold-atom experiments with nonequilibrium dynamics.


It is exciting to contemplate the ways in which this setup can enable new explorations.
For example, the Fermi gas would be useful in damping out any excitations that are caused by transferring the bosons into an optical lattice \cite{PhysRevLett.96.180402, PhysRevLett.97.220403}.  Alternatively,
if one stirred the Fermi gas, the dissipative forces would bring the BEC into equilibrium in the rotating frame -- producing a vortex lattice.
More generally, engineered dissipation is a powerful tool, which we are just beginning to explore.








\begin{acknowledgments}
We thank the Chin group for discussions about the experiments.
This work was supported by the NSF Grant PHY-1806357 and the ARO-MURI Non-Equilibrium Many-body Dynamics Grant W9111NF-14-1-0003.
\end{acknowledgments}

\appendix

\section{Simulation of Boson and Fermion Density Profiles} \label{densitysimulation}
Here we give details about how we numerically solve
Eq.~(\ref{bosondensity}) and Eq.~(\ref{fermiondensity}). We work in cylindrical coordinates, with the symmetry axis labeled as $\bf \hat x$.
We choose a simulation box that
is sufficiently large to enclose the entire Fermi cloud.
We then set up a 100$\times$100 grid, and initialize the boson wavefunction $\psi$ and the Fermi density $n_F$ to the values they would take in the absence of $g_{BF}$.  

\begin{figure*}
\includegraphics[width=2.2\columnwidth]{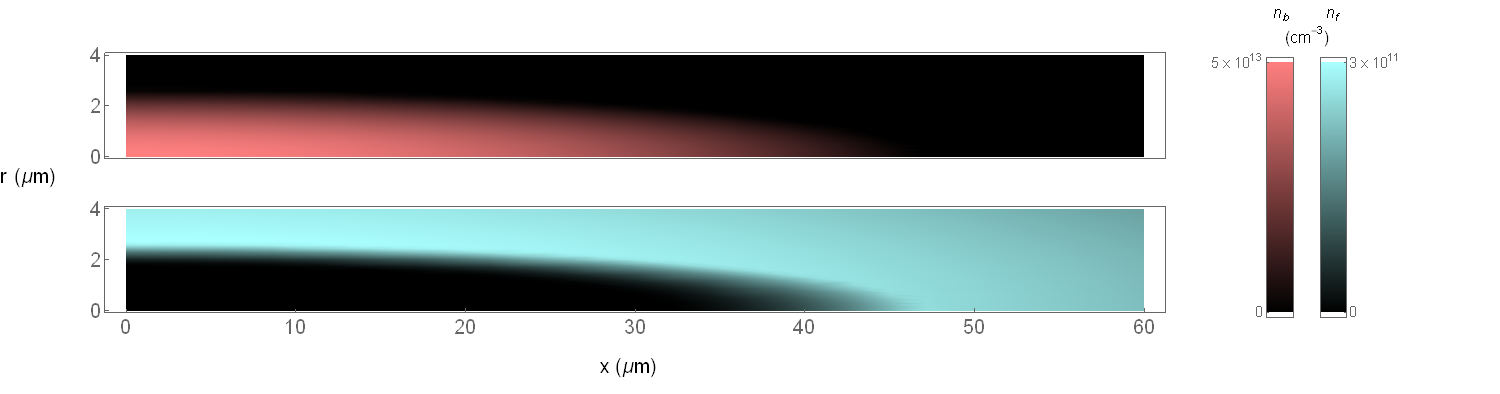}
\caption{(Color online) Simulated density profiles of bosons (top, red) and fermions (bottom, blue) for $a_{BF} = 340 a_0 $.  The long axis is $x$, and the perpendicular direction is $r$. Lighter colors correspond to higher densities.  Note the non-monotonic behavior of the Fermi density, as schematically shown in Fig.~\ref{buoyancydiagram}.}
\label{kin-kout}
\end{figure*}

First fixing $n_F$, we optimize $\psi$ using gradient descents, minimizing the  energy of the system (this has been done before by \cite{Ufrecht_2017, capuzzi2004}).  We then update $n_F$ via Eq.~(\ref{fermiondensity}) -- adjusting $\mu_F$ to keep $N_F$ fixed.  We continue to cycle through these two steps until convergence.  Typically about 10 iterations are needed.  Fig~\ref{kin-kout} shows a typical result for a moderately strong repulsive Bose-Fermi interaction strength.  The Fermi density vanishes in the central region, rises as one approaches the edge of the Bose cloud, then falls again as one moves towards the edge of the trap.  The Bose density simply falls monotonically.



\section{Second order expansion of $\lambda$}\label{higherorderlambda}
Here we continue the argument from Sec.~\ref{mfc}, and show that there are no mean-field contributions to the damping at second order in $\lambda$.

Recall, we have introduced the perturbative expansion
\begin{align}
    \boldsymbol{k}(t) &=  \boldsymbol{k}_i + g_{BF} \boldsymbol{k}^{(1)}(t) + \frac{g_{BF}^2}{2} \boldsymbol{k}^{(2)}(t) + \hdots\\
    \boldsymbol{r}(t) &=  \boldsymbol{r}^{(0)}(t) + g_{BF} \boldsymbol{r}^{(1)}(t) + \frac{g_{BF}^2}{2} \boldsymbol{r}^{(2)}(t) + \hdots
\end{align}
Substituting these into Newton's Laws yields
\begin{eqnarray}
\frac{d k_\mu}{dt}&=&
-\frac{1}{\hbar} \nabla_\mu V({\bf r})\\ \nonumber
&=& -\frac{1}{\hbar}\nabla_\mu V({\bf r}^{(0)})
-\sum_\nu g_{BF}\frac{1}{\hbar} r_\nu^{(1)} \nabla_\mu\nabla_\nu V({\bf r}^{(0)})+\cdots
\end{eqnarray}
The potential $V$ is proportional to $g_{BF}$, and hence
\begin{eqnarray}
k_\mu^{(2)} &=& -\frac{1}{m_F} \int_0^t dt_1 \,\sum_\nu r_\nu^{(1)}\nabla_\mu\nabla_\nu V\left(
    \boldsymbol{r}^{(0)}
    \right)\\
    &=& 
    \int_0^t dt_1\int_0^{t_1} dt_2 \int_0^{t_2} dt_3\,
\nabla_\mu \Lambda({\bf r}^{(0)}(t_3))
\end{eqnarray}
where we used 
\begin{eqnarray}\nonumber
\boldsymbol{r}^{(1)}(t_1) &=& -\frac{1}{m_F} \int_0^{t_1} dt_2 \int_0^{t_2} dt_3 \, \nabla V\left(
    \boldsymbol{r}^{(0)}(t_3)\right).
\end{eqnarray}
to get
\begin{eqnarray}
\Lambda({\bf r})&=&
\frac{1}{\hbar m_F} |\nabla V(\bf r)|^2.
\end{eqnarray}
The second order contribution to $\lambda$ is then 
\begin{eqnarray}
\lambda^{(2)} &\propto&\int_{\hspace{-2mm}\raisebox{-2.5mm}{\mbox{$\scriptstyle |\bm{k_i}|=k_{\rm fermi}$}}}\hspace{-1.1cm}\frac{d^3\bm{k_i}}{(2\pi)^3}(\hat{\bm{v}} \cdot \bm{k_i}) 
\int d^2 \bm{r_\perp} F
\end{eqnarray}
with
\begin{eqnarray}
F&=&
 \int_0^T dt_1 \int_0^{t_1} dt_2 \int_0^{t_2} dt_3 \, \hat{\boldsymbol{v}}\cdot\nabla\Lambda({\bf r}^{(0)}(t_3))
\end{eqnarray}
We swap the order of integrals, to make the inner integral
\begin{equation}
I= \int  d^2 \bm{r_\perp} \nabla\Lambda.
\end{equation}
The gradient can be broken into parts parallel and perpendicular to ${\bf k_i}$: $\nabla=\nabla_\perp+\nabla_\parallel$.  The integral of the perpendicular part vanishes by Stoke's theorem.  The parallel part can also be shown to vanish by converting the $t$ integrals into spatial integrals, and using the Fundamental Theorem of Calculus.

 \section{Lensing model of frequency shift}\label{montecarlofrequency}

Here we give an alternative model for the frequency shift of the Boson dipole mode, which is based directly upon kinetic theory.  In particular, we will show that if the Boson cloud is immersed in a Fermi cloud of non-uniform density, it will experience a force. 
 Following our treatment of the drag forces in Sec.~\ref{drag}, we will calculate 
 this force
 as the third law pair to the forces experienced by individual fermions incident on the cloud.

 As illustrated in Fig.~\ref{trajectoryschematic}, when $g_{BF}>0$, fermions which hit the right side of the Bose cloud tend to bend to the right.  Conversely, fermions which hit on the left side tend to bend to the left.  Thus if there are more fermions on the right then the bosons will experience a net force to the left.  The opposite happens when $g_{BF}<0$.

 This approach is another way of thinking of buoyancy:  Buoyancy forces can be calculated by adding up the forces from fluid pressure.  When the pressure is non-uniform (as is the case with a density gradient) there will be a net force.

Within error bars, this kinetic approach agrees with the buoyancy model.
 
 

 \subsection{Linearizing}
 
We linearize the fermion trap potential around the center of the Bose cloud, so that the potential felt by the fermions is
 \begin{equation}
V_F({\bf r})=g_{BF} n_B({\bf r})+ F x
 \end{equation}
 where $n_B$ is the equilibrium configuration of the bosons, calculated in Appendix~\ref{densitysimulation}. The force  $F$ comes from the external potential, leading us to take:
 \begin{equation}
F=m_F\omega_F^2 X_B,
 \end{equation}
 where $X_B$, the center of mass of the bosons, is considered fixed during our calculation of the fermion trajectories.  
 
 Similar to the argument in Sec.~\ref{kinetics}, we will calculate the momentum transferred from the bosons to the fermions in a time $\Delta t$, and to linear order in $F$ will have $\Delta p=\chi F \Delta  t$.  The dimensionless proportionality constant  $\chi$ will be a function of scattering lengths.
Including the drag force, the equation of motion for the boson cloud is
 \begin{align}
     \ddot{X_B} =  - \omega_B^2 X_B - 2\Gamma \dot{X_B} + \frac{\chi m_F^2 \omega_F^2 X_B}{M_B}  
 \end{align}
 The frequency shift of Boson cloud oscillations is therefore proportional to $\chi$:
 \begin{align} \label{freqshiftexpression}
     \delta \omega \approx - \frac{ m_F^2 \omega_F^2 }{2 \omega_B M_B }\chi.
 \end{align}

Consider a fermion which is initially at position $\bf R$ with momentum $\hbar {\bf k_i}$.  In a stochastic process, it will interact with the Boson cloud.  We denote the average momentum transfer as
$G_{{\bf k_i},{\bf R}}^{(F)}$.   The $F$ dependence comes from the fact that the external force causes the fermion's trajectory to curve.   We can find the total momentum transfered to the Boson cloud in time $\Delta t$ as
\begin{equation}
\Delta p=
\int \frac{d^3 k_i}{(2\pi)^3}
\oint_\Omega d^2 {\bf R}\, \frac{\hbar |{\bf k_i}|}{m_F} 
G_{k_i,R}^{(F)} \, f\left(\frac{k_i^2}{2m_F}+Fx\right)\, \Delta t
\end{equation}
The launch points lie on a closed surface $\Omega$, which encloses the boson cloud -- and the  result should not depend on the choice of surface. In our simulation, we choose to use a cylindrical ``box'' that is large enough to enclose the boson cloud fully.
The unit vector $\hat n$ points perpendicular to this surface.
As used elsewhere, $f(E)=\theta(\mu_F-E)$ is the  Fermi function.  The combination $\left({\hbar |{\bf k_i}|}/{m_F}\right) \Delta t$ is a geometric factor which gives the volume of particles which pass through $\Omega$ in time $\Delta t$.

We linearize for small $F$ to  arrive at $\chi=\chi_1+\chi_2$, where:
\begin{align}\label{chi}
\chi_1&=
\int \frac{d^3 k_i}{(2\pi)^3}
\oint_\Omega d^2 {\bf R}\, \frac{\hbar |{\bf k_i}|}{m_F} 
G_{k_i,R}^{(0)}\, x\, \delta\left(\frac{k_i^2}{2m_F}-\mu\right) \\\label{chi2}
\chi_2&=\int \frac{d^3 k_i}{(2\pi)^3}
\oint_\Omega d^2 {\bf R}\, \frac{\hbar|{\bf k_i}|}{m_F} \frac{\partial G_{k_i,R}^{(F)}}{\partial F}
 f\left(\frac{k_i^2}{2m_F}+Fx\right).
\end{align}
Here $\chi_1$ involves particles at the Fermi surface, and is related to the effect of $F$ on the distribution function.  Conversely, $\chi_2$ involves particles within the Fermi sea, and is related to the effect of $F$ on their trajectories.
Note that each component above depends on $x$, the x-coordinate of each fermion's launch point, and both $\chi_1$ and $\chi_2$ depend on the choice of launch surface $\Omega$. However, this dependence should cancel out in the final result for $\chi$.
  
\begin{figure}[tbp]
 {\includegraphics[width=\columnwidth]{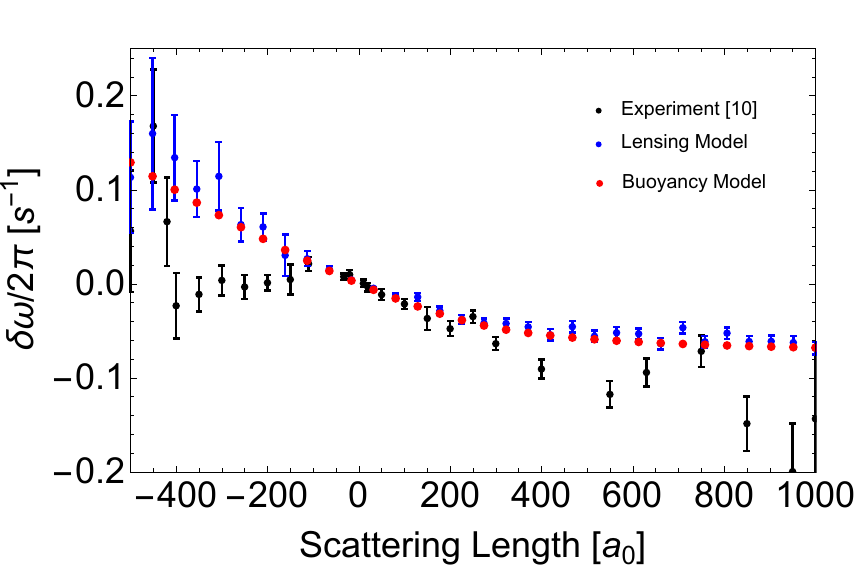}}
      \caption{(Color Online) Dipole mode frequency shift plot identical to Fig.~\ref{freqshiftcomparison}, but with the lensing model predictions added as blue points. Each point represents 10000 samples of fermion trajectories, with error bars representing the
      standard error of the mean. The lensing model gives predictions that are consistent with the buoyancy model.}
      \label{lensingfigure}
  \end{figure}

 \subsection{Calculation of  $G_{k_i,R}^{(0)}$ and $\partial G_{k_i,R}^{(F)}/\partial F$}

Consider the trajectory of a fermion moving through the Boson cloud. At a time $t$, the fermion is specified by a position ${\bf r}(t)$ and ${\bf p}(t)$. Further, let $t=t_i$ be the time at the start of the trajectory, and $t=t_f$ be the end time of the trajectory in the simulation. To first order in $F$, we can linearize both variables:
\begin{align}
    {\bf r}(t) = {\bf r}_0(t) + F  {\bf \delta r}(t)+{\cal O}(F^2)\\
    {\bf p}(t) = {\bf p}_0(t) + F  {\bf \delta p}(t)+{\cal O}(F^2).
\end{align}
That is ${\bf r}_0(t)$ and ${\bf p}_0(t)$ are the trajectories in the absence of $F$, while $F  {\bf \delta r}(t)$ and $F  {\bf \delta p}(t)$ are the first order corrections.  We substitute these into Newton's laws:
$\partial{\bf r}/\partial t = {\bf p}/m,$ 
$\partial{\bf p}/\partial t = -\nabla V_F({\bf r}) = -g_{BF} \nabla n_B({\bf r}) - F \hat{x}$
and expanding to linear order in $F$:
\begin{align}
    \dol{{\bf p}_0}{t} + & F\dol{\delta{\bf p}}{t}\\\nonumber&= -g_{BF} \left(\nabla n_B({\bf r}_0) + F (\delta {\bf r}\cdot \nabla) \nabla n_B({\bf r}_0) \right) - F \hat{x}
\end{align}
Separately equating terms which are independent of F, and those which are of first order in F we get:
\begin{align}\label{ptimestep}
    {\bf p}_0(t+\delta t) &= {\bf p}_0(t) -  g_{BF} \, \delta t \, \nabla  n_B({\bf r}_0)\nonumber\\
    \delta{\bf p}(t+\delta t) &= \delta{\bf p}(t) - \delta t \left(  g_{BF}(\delta {\bf r}\cdot \nabla) \nabla n_B({\bf r}_0) +  \hat{x}\right)\nonumber \\
    {\bf r}_0(t+\delta t) &= {\bf r}_0(t) + {\delta t}\,{\bf p}_0(t+\delta t) /m_F \nonumber \\
     \delta{\bf r}(t+\delta t) &= \delta{\bf r}(t) + {\delta t}\, \delta{\bf p}(t+\delta t)/m_F.
\end{align}
As in section~\ref{montecarlo}, use these stepping rules to evolve the fermion trajectory, including scattering through a stochastic process.  We then calculate
\begin{align}
    G_{k_i,R}^{(0)} = {\bf p}_0(t_f) - {\bf p}_0(t_i)\\
    \dol{ G_{k_i,R}^{(F)}}{F} = {\bf \delta p}(t_f) - t_f \hat{x}
\end{align}
We evaluate Eq.~(\ref{chi}) and (\ref{chi2}) as a Monte Carlo sum:
 \begin{align}
 \chi_1 &= \frac{\hbar k_{\rm fermi}^3 }{2\pi^2  m_F}  \frac{1}{N}\sum_{i=1}^N  A_{i} x_i \,  G_{k_i,R}^{(0)} \\
    \chi_2 &= \frac{\hbar k_{\rm fermi} }{2\pi^2  m_F}  \frac{1}{N}\sum_{i=1}^N  A_{i} |{\bf k_i}|^3 \dol{ G_{k_i,R}^{(F)}}{F}
 \end{align}
where $i$ denotes the sample index,  $N$ is the total number of samples, $x_i$ is the x-coordinate of each sample's launch point, and $A_i=4 c \xi(\theta_i,r_{1,i})$ is the area of the aperture as described in Sec.~\ref{montecarlo}. In choosing the sample parameters, we again  choose $\cos\theta$ and $r_1$  uniformly in $[-1,1]$ and $[-c,c]$ respectively, and then choose $r_2$ uniformly in $[-\xi(\theta,r_1),\xi(\theta,r_1)]$.  We use separate trajectories for calculating 
$\chi_1$ and $\chi_2$ sums. For $\chi_1$, $|k_i| = k_{\rm fermi}$ for all samples; while for $\chi_2$, $|k_i|$ is chosen uniformly in $[0,k_{\rm fermi}]$. $G_{k_i,R}^{(0)}$ and $\partial G_{k_i,R}^{(F)}/\partial F$ are then calculated using the expression in Eq.~(\ref{ptimestep}).

 Finally, we calculate the frequency shift via Eq.~(\ref{freqshiftexpression}). Fig.~\ref{lensingfigure} compares the results of this lensing model to the Buoyancy model. They agree within error bars, and both deviate from the experimental data in the same way. 

\bibliography{aps}

\end{document}